\documentclass[12pt,preprint]{aastex}
\usepackage{natbib}
\def\gtabouteq{\,\hbox{\raise 0.5 ex \hbox{$>$}\kern-.77em
                    \lower 0.5 ex \hbox{$\sim$}$\,$}}
\def\ltabouteq{\,\hbox{\raise 0.5 ex \hbox{$<$}\kern-.77em
                     \lower 0.5 ex \hbox{$\sim$}$\,$}}

\slugcomment{Submitted to ApJ}

\shorttitle{Extrasolar planet inclinations}
\shortauthors{Thommes and Lissauer}

\begin{document}
\title{Resonant inclination excitation of migrating giant planets}
\author{Edward W. Thommes}
\affil{Astronomy Department, University of California, Berkeley, CA
94720}

\and

\author{Jack J. Lissauer} \affil{Space Science Division, MS 245-3, NASA-Ames Research Center, Moffett Field, CA 94035}

\begin{abstract}
The observed orbits of extrasolar planets suggest that many giant
planets migrate a considerable distance towards their parent star as a
result of interactions with the protoplanetary disk, and that some of
these planets become trapped in eccentricity-exciting mean motion
resonances with one another during this migration.  Using
three-dimensional numerical simulations, we find that as long as the
timescale for damping of the planets' eccentricities by the disk is
close to or longer than the disk-induced migration timescale, and the
outer planet is more than half the mass of the inner, resonant
inclination excitation will also occur.  Neither the addition of a
(simple, fixed) disk potential, nor the introduction of a massive
inner planetary system, inhibit entry into the inclination resonance.
Therefore, such a mechanism may not be uncommon in the early evolution
of a planetary system, and a significant fraction of exoplanetary
systems may turn out to be non-coplanar.

\end{abstract}

\section{Introduction}
The orbital migration of bodies locked in mean-motion resonances
is a major recurring theme in the early history of the Solar
System. A review of resonance dynamics in the Solar System is
given by \cite{1976ARA&A..14..215P}.  \citet*{1965MNRAS.130..159G} showed that
tidally-induced orbital migration is likely responsible for most
of the mean-motion commensurabilities among the satellites of Jupiter and
Saturn.  In the Uranian satellite system, tidally-induced
convergent migration of the moons Miranda and Umbriel could have
led to capture into, and subsequent escape from, inclination-type
resonances, thus producing Miranda's anomalously high orbital
inclination of $4.3\degr$ \citep*{1989Icar...78...63T, 1990Icar...85..444M}.

In the case of the outer Solar System, \cite{1998LPI....29.1476M} and
\cite{2000AJ....120.2695G} showed that bodies entrained in the 2:3
exterior mean-motion resonance of an outward-migrating Neptune can
also encounter the Kozai resonance, as well as secular and mean-motion
inclination resonances.  These resonances increase inclination, and
this may be the mechanism by which Pluto and the Kuiper belt objects
presently sharing the 2:3 resonance (plutinos) attained their high
inclinations.

\citet{2001A&A...374.1092S} and \cite{2002ApJ...567..596L} performed
in-depth investigations of resonant capture and migration in the
context of the GJ 876 planetary system.  They found that convergent
orbital migration of planets typically leads to their capture into the
2:1 mean-motion resonance.  The resonant lock thus produced is
characterized by the libration of the first-order resonance angles
(cosine arguments in the expansion of the disturbing function; see for
example \cite{1999..Murray..Dermott..book}),
\begin{eqnarray}
\theta_{e_1} \equiv \lambda_1- 2 \lambda_2 + \tilde{\omega}_1&\mbox{and}
\nonumber \\
\theta_{e_2} \equiv \lambda_1- 2 \lambda_2 + \tilde{\omega}_2&\,
\label{res_angles_1}
\end{eqnarray}
about zero degrees, where $\lambda_i$ is the $i$th planet's longitude,
$\tilde{\omega}_i$ is its longitude of pericenter, and $i=1, 2$ refer
to the inner and outer planet, respectively.  Since
$\theta_{e_1}-\theta_{e_2}=\tilde{\omega}_1 - \tilde{\omega}_2$,
the zeroth-order secular resonance
angle
\begin{equation}
\theta_{\tilde{\omega}}\equiv\tilde{\omega}_1 -
\tilde{\omega}_2
\label{apsidal_res_angle}
\end{equation}
also librates about zero;
in other words, the orbits of the two planets librate about apsidal
alignment.

The analysis of \cite{2002ApJ...567..596L} is restricted to the planar
case, thus the effects of migration in resonance on orbital
inclinations are not considered.  Starting with slightly non-coplanar
orbits, we initially see the same behaviour as reported by
\cite{2002ApJ...567..596L}: In the absence of any damping mechanism,
the eccentricities of both bodies grow.  However, we find that once
the inner planet's eccentricity is high enough, the system will also
enter an inclination-type resonance, which induces rapid growth in the
inclinations of both planets.

In this article, we report the results of three-dimensional
simulations of migrating planets in resonance.  We give a summary of
the underlying physical picture in Section \ref{phys picture}.  The
methods that we use, as well as initial conditions, are presented in
Section \ref{numerical methods}.  Baseline cases with two equal mass
planets are discussed in Section \ref{baseline}.  We extend our
analysis to planets of unequal masses in Section \ref{mass ratio},
impose eccentricity damping in Section \ref{damping}, include the
gravitational potential of the disk in Section \ref{disk potential},
and add four Uranus/Neptune mass inner planets in Section \ref{addl
planets}.  We conclude in Section \ref{discussion} with a summary of
our results and a discussion of their implications.

\section{The physical picture}
\label{phys picture}
Gas giant planets must form in a protoplanetary disk which has not yet
lost its gas; this constrains the time of formation to within,
approximately, the first $10^7$ years of the disk's lifetime
\citep*{1990csss....6..275S,2001ApJ...553L.153H}.  If gas giants form
by the accretion of a massive atmosphere onto a $\sim$ 10 M$_{\oplus}$
solid core (\citealp*{1978PThPh..60..699M};
\citealp{1996Icar..124...62P}; \citealp{2000Icar..143....2B}),
then the most favourable location for such large cores to form by
planetesimal accretion is in an annulus several AU wide beyond the
ice-condensation line, roughly the location of Jupiter and Saturn in
our own Solar System \citep*{2003Icarus...tdl...oligarchy}.  A gas
giant-sized body (mass of order $10^2$ M$_{\oplus}$) will open a gap
in the gas disk.  Its orbital evolution will subsequently be locked in
to the viscous evolution of the gas disk, as long as the disk mass is
large compared to that of the planet.  This is commonly referred to as
Type II migration \citep{1997Icar..126..261W}; see
\citet{2000prpl.conf.1111L} or \citet*{thommes_lissauer_03} for a
review.  In this picture, significant Type II migration probably did
not occur in our Solar System; however, this mechanism is likely to
have played a significant role for those extrasolar giant planets with
orbital semimajor axes of less than a few AU.  For a review of
discovered extrasolar planets and the statistics of their orbits, see
\cite{2003asp...cumming}.

If a disk forms two gas giants, the annulus of gas between their
individual gaps may be relatively rapidly depleted, so that both
planets end up sharing a common gap
\citep{2000MNRAS.313L..47K,2000ApJ...540.1091B} Also, if the viscosity
is higher and thus the accretion of the disk onto the star faster at
smaller radii, as in the model of \citet*{1996ApJ...457..355G}, the
inner disk may be depleted on a timescale short compared to the
migration timescale of the planets.

\section{Numerical methods}
\label{numerical methods}

We use a variant of the SyMBA integrator \citep*{1998AJ....116.2067D},
which has the desirable properties of the mixed-variable symplectic
(MVS) mapping of \citet*{1991AJ....102.1528W}, and in addition handles
close encounters among objects by employing a variable timestep
technique.  We impose an inward migration on the outer planet, in a
way which simulates interaction with a viscously evolving outer gas
disk: We define an inner radius $r_{\rm disk}$ to the gas disk; this
radius is made to change at a rate $\dot{r}_{\rm disk}$.  The
semimajor axis of the outer planet, $a_2$, is then made to change based
on its location relative to the gap edge:
\begin{equation}
\dot{a}_2 = \left \{ \begin{array}{l l}
-S (a_2-r_{\rm disk}), & a_2 > r_{\rm disk}\\
0, & a_2<r_{\rm disk}
\end{array} \right .
\label{gap prescription}
\end{equation}
where $S$ is large and positive, so that the magnitude of
$\dot{a_2}$ increases steeply outside of $r_{\rm disk}$.  In
this way, a planet with $a_2 > r_{\rm disk}$ is always
constrained to move inward at a mean rate $\dot{r}_{\rm disk}$; if the
torque required to maintain this migration rate changes (as happens
when the planet resonantly captures an interior body which it then has
to push along), the planet's position relative to the gap edge simply
re-adjusts until its orbital decay rate again matches $\dot{r}_{\rm
disk}$.  It should be noted that, strictly speaking, $r_{\rm disk}$
corresponds not to the edge of the disk itself but rather to the
location outwards of which a planet is subject to a nonzero disk
torque.  

The speed with which the disk edge moves inward corresponds to the
radial accretion speed of the disk.  This is given by
\begin{equation}
\dot{r}_{\rm accr} \sim C_1 \alpha (r \Omega) (c/r \Omega)^2,
\label{viscous accretion rate}
\end{equation}
where $\Omega$ is the Keplerian angular frequency, $c$ is the gas
sound speed, $C_1$ is a constant of order unity, and $\alpha$ is the
Shakura-Sunyaev viscosity parameter: $\nu \sim \alpha c^2/\Omega$
\citep*[e.g.,][]{1997Icar..126..261W}.  The factor $(c/r \Omega)$ is
also the aspect ratio $h$ of the disk, where $h \equiv H/r$; $H$ is
the disk scale height at radius $r$. For simplicity we adopt the
(flared) aspect ratio of the \cite{1981PThPh..70...35H} nebula model,
which has $h \propto r^{1/4}$; \cite{1997ApJ...490..368C} obtain a
similar $r$-dependence, $h \propto r^{2/7}$, for $r \ltabouteq$ 80 AU.
We thus have a disk accretion speed $\dot{r}_{\rm accr}$ which does not
vary with $r$, assuming $\alpha$ is constant.  Using $h=0.05
(r/\mbox{AU})^{1/4}$, the accretion speed is $\dot{r}_{\rm accr} \sim
10^{-2} \alpha$ AU/year.  The $\alpha$ viscosity of a protoplanetary
disk is estimated to be $10^{-4}$ to $10^{-2}$
\citep[e.g.,][]{1987Icar...69..387C,1993Icar..106...59D}.  Adopting
$\alpha = 10^{-3}$, we have $\dot{r}_{\rm accr} \sim 10^{-5}$ AU/year.

It should be emphasized that migration is imposed on the outer planet
as per Eq. \ref{gap prescription} regardless of the eccentricity of
the planet, or its inclination relative to the plane of the disk
(which is the planet's original orbital plane).  Hydrodynamic
simulations of the interaction of a significantly eccentric and/or
inclined gap-opening planet with a disk have, to our knowledge, not
yet been performed, so we operate under the simple assumption that, no
matter how the planet's orbit evolves, its semimajor axis remains
locked relative to the inner edge of the 
disk.  However, at sufficiently high eccentricity and/or inclination,
a planet will very likely no longer be an effective barrier to the
disk, so that this assumption would no longer be valid.  We will
discuss implications in Section \ref{discussion}.

The shortest base timestep used in the simulations is $10^{-3}$ years,
and integration is stopped when a body comes within 0.1 AU of the
star.  Beyond this point, the timestep is deemed too long, and thus the
azimuthal motion of the inner planet during one timestep too
large, to accurately resolve the interactions between the two
planets.

In all of the simulations below, we start with a system of two planets
on nearly circular orbits about a one solar mass (M$_{\odot}$) star,
the inner at 5 AU, and the outer at 8.5 AU.  Initial eccentricities
and inclinations are of order $10^{-3}$ and $(10^{-2})\degr$,
respectively.

\section{The baseline case}
\label{baseline}

\begin{figure}
\begin{center}
\includegraphics[width=6.0in]{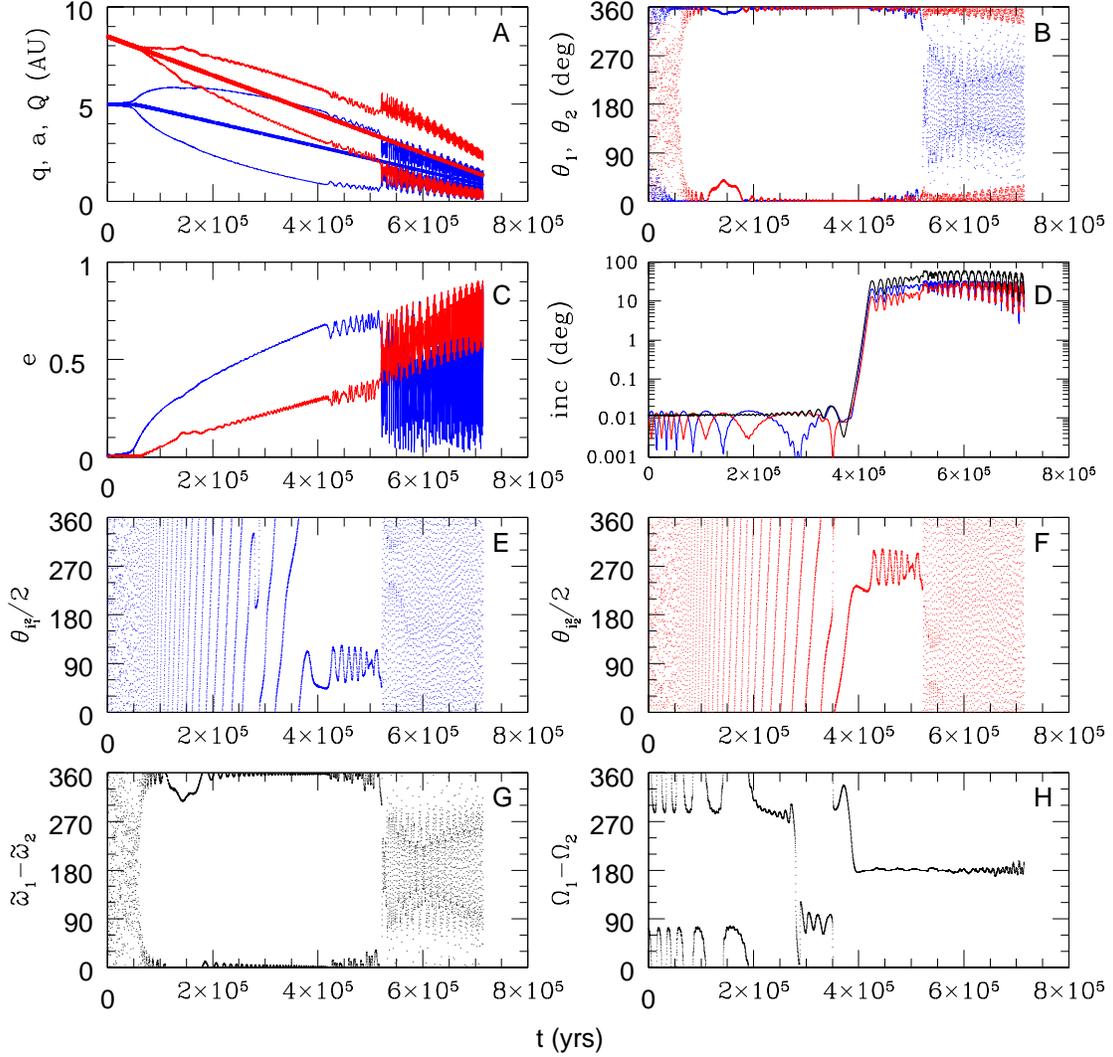}
\caption{Evolution of the two planets in the baseline run, in which an
inward migration rate of $\dot{a}=10^{-5}$ AU/year is imposed on the
outer one. Both planets are one Jupiter mass (1 M$_J$); quantities
related to the inner (Planet 1) are plotted in blue, and in
red for the outer (Planet 2). The panels show: A. Semimajor axis,
periastron distance and apastron distance; B. The lowest-order
eccentricity-type resonance angles $\theta_{e_1}$ (blue) and
$\theta_{e_2}$ (red); C. Eccentricities of both planets; D.
Inclinations of both planets, as well as (black) the mutual
inclination; E., F. $\theta_{i_1^2}/2$ and $\theta_{i_2^2}/2$ where
$\theta_{i_1^2}$ and $\theta_{i_2^2}$ are the lowest-order
inclination-type resonance angles; G. The angle between the planets'
longitudes of periastron; H. The angle between the planets' lines of
nodes.} \label{baseline1}
\end{center}
\end{figure}

We begin with a system consisting of two Jupiter-mass planets.
Several runs are performed, with differing initial orbital phases, but
the outcomes are very similar.  Fig. \ref{baseline1} shows one
example.  The outer planet migrates inward until it and the inner
planet are mutually captured into the 2:1 mean-motion resonance, with
both first-order eccentricity-type resonance angles, $\theta_{e_1}$
and $\theta_{e_2}$ (Eq. \ref{res_angles_1}), librating about zero,
which means that conjunctions are occurring when both planets are at
pericenter, and thus that the orbits are also librating about apsidal
alignment.  This behaviour is to be expected even for planets that
start with a larger separation in orbital periods;
\citet{2002ApJ...567..596L} find that as long as $\dot{a_2} \la
10^{-3}(a/{\rm AU})^{-1/2}$, capture into the 2:1 resonance occurs
with certainty if the eccentricities of both planets are low ($\la
10^{-1}$), whereas larger eccentricities are required for a nonzero
capture probability into more distant, higher-order resonances which
would be encountered first (such as 3:1).

The eccentricities of both planets increase as they migrate in
resonance, which can be understood in terms of energy ($E$) and
angular momentum ($L$) exchange between the disk and the system of
resonantly-locked planets \citep*{1984Icar...58..159L}: In order to
change the semimajor axis of a circular orbit while keeping it
circular, $E$ and $L$ have to be changed in the right ratio:
$\dot{E}/\dot{L} = dE/dL = \sqrt{G M_*/a^3}$, which is just the
orbital angular velocity, $n(a)$.  But $n(a)$ is a decreasing function
of semimajor axis.  Thus, even though we assume a disk torque that
changes the orbit of the initially circular outer planet in such a way
that $\dot{E}/\dot{L}=n(a_2)$, this ratio is too low for the inner
planet.  Since $\dot{E}$ and $\dot{L}$ are negative, the two-planet
system as a whole acquires an angular momentum deficit and must
respond by increasing the eccentricity of one or both of its
components.  The rate of eccentricity growth can be reduced by making
the applied $\dot{E}/\dot{L}$ {\it higher} than what would be required
to keep the outer planet by itself circular.  This would amount to
applying eccentricity damping to the outer planet.  The evolution of
the system with eccentricity damping by the disk is investigated in
Section \ref{damping}.

In Fig. \ref{baseline1}E and F, we also plot the resonance angles for
the 4:2 inclination-type mean motion resonance.  This is a
second-order resonance (which is the lowest possible order for an
inclination resonance); the associated terms in the expansion of the
disturbing function are of order $i^2$.  These angles are defines as
\begin{equation}
\theta_{i_1^2} = 2 \lambda_1 - 4 \lambda_2 + 2 \Omega_1
\label{theta_i1sq}
\end{equation}
and
\begin{equation}
\theta_{i_2^2} = 2 \lambda_1 - 4 \lambda_2 + 2 \Omega_2,
\label{theta_i2sq}
\end{equation}
with $\Omega_1$, $\Omega_2$ being the longitudes of the ascending
nodes of the inner and outer planet, measured relative to an arbitrary
but fixed direction in the inertial frame, and in the plane of the
(notional) protoplanetary disk, which is, within an inclination of
order $(10^{-2})\degr$, also the initial plane of both planets' orbits.

Both angles are initially circulating, but $\dot{\theta_{i_1^2}}/2$
and $\dot{\theta_{i_2^2}}/2$ decrease over time until, at about $4
\times 10^5$ years, $\theta_{i_1^2}/2$ starts librating about about
$90\degr$, while simultaneously $\theta_{i_2^2}/2$ starts librating
about $270\degr$.  This means that ${\theta_{i_1^2}}$ and
${\theta_{i_1^2}}$ are both librating about $180\degr$.  Fig.
\ref{baseline1}D shows that initially, the inclinations of both bodies
remain at $\la 0.01\degr$, oscillating but undergoing no net growth.
However, as the system enters the inclination resonances, the
inclinations of both bodies begin to grow rapidly to $\sim 10\degr$
(the outer planet) and $\sim 20\degr$ (the inner one) in just tens of
thousands of years.  Thereafter, the inclinations continue to grow
more slowly.

This simultaneous libration of $\theta_{i_1^2}$ and
$\theta_{i_2^2}$ also implies the libration of the ``mixed''
resonance angle,
\begin{equation}
\theta_{i_1 i_2} = 2 \lambda_1 - 4 \lambda_2 + \Omega_1 + \Omega_2
 = \theta_{i_1^2}/2+\theta_{i_2^2}/2
\label{theta_i1i2}
\end{equation}
about $0\degr$, as well as libration of the zeroth-order secular
resonance angle,
\begin{equation}
\theta_{\Omega} \equiv \Omega_1 -
\Omega_2=\theta_{i_1^2}/2-\theta_{i_2^2}/2
 \label{theta_Omega}
\end{equation}
about $180\degr$.  The latter means that the two orbits are
librating about anti-alignment of their lines of nodes.  Fig.
\ref{baseline1}H shows that the libration amplitude is only a few
degrees.

Furthermore, simultaneous libration of $\theta_{e_1}$,
$\theta_{e_2}$, and $\theta_{\Omega}$ also implies the libration
of the resonance angles
\begin{equation}
\theta_{e_1i_1i_2} \equiv \lambda_1 - 2\lambda_2 +\tilde{\omega}_1
\pm(\Omega_1-\Omega_2) = \theta_{e_1} \pm \theta_{\Omega}
\label{e1i1i2}
\end{equation}
and
\begin{equation}
\theta_{e_2i_1i_2} \equiv \lambda_1 - 2 \lambda_2+\tilde{\omega}_2
\pm(\Omega_1-\Omega_2) = \theta_{e_2} \pm \theta_{\Omega}.
\label{e2i1i2}
\end{equation}

\begin{figure}
\begin{center}
\includegraphics[width=6.0in]{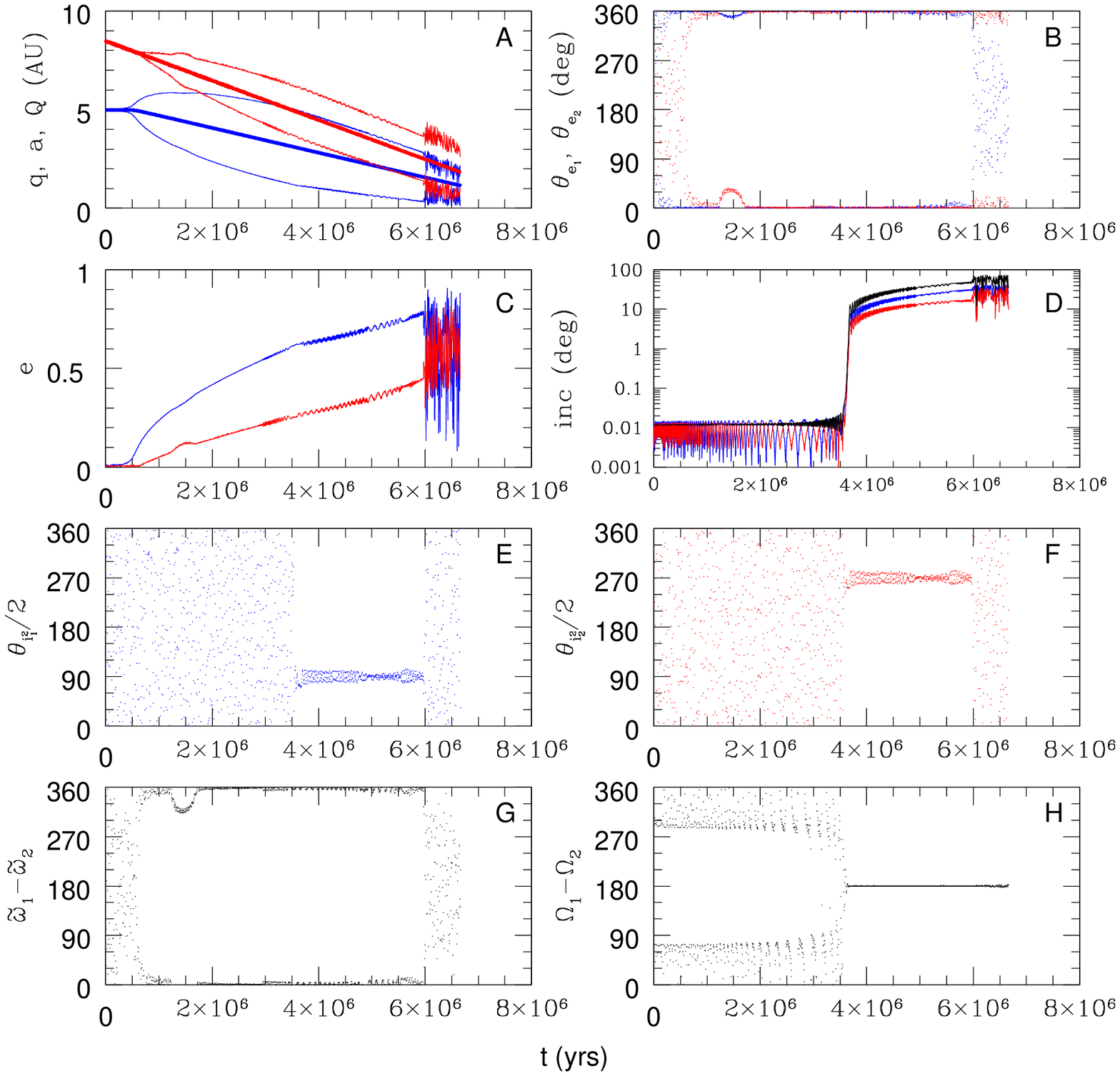}
\caption{Evolution of the planets when the migration rate imposed
on the outer planet is $10^{-6}$ AU/year, one tenth the rate in the
baseline run (Fig. \ref{baseline1}).  All panels are as in Fig.
\ref{baseline1}. } \label{slow_migr}
\end{center}
\end{figure}

The near-perfect anti-alignment of the lines of nodes can be
understood in terms of the conservation of that component of the
system's angular momentum which lies in the original orbital plane.
Since this is initially very near to zero, the in-plane components of
the two orbits' angular momentum vectors have to be pointing in very
nearly opposite directions once the inclinations get large, otherwise
they cannot sum to (almost) zero.  This also means that the {\it
relative} inclination of the two planets' orbital planes (see
\ref{baseline1}D) is the sum of their individual inclinations.  Thus
the relative inclination of the two planets very quickly goes up to
$30 \degr$.

At $5.2 \times 10^5$ years, the system evolves out of both
inclination-type resonances. There is an abrupt increase in
inclinations at this time, but thereafter, as is to be expected, there
is no further net inclination increase. At the same time, the inner
planet's eccentricity-type resonance angle, $\theta_{e_1}$, switches
from libration about $0\degr$ to ``near-libration'' about $180\degr$:
The angle librates about $180\degr$ for a few cycles, then makes a
complete revolution, librates for another few cycles, and so on. From
a perturbative analysis point of view, this means that the
contribution from the corresponding term in the expansion of the
disturbing function continues to be important; its cosine argument
$\theta_{e_1}$ spends most of its time around $180 \degr$ so the term
does not average to zero over time.  An interesting point is that in a
completely planar system, this transition in libration center is
inevitably cut short by the system becoming unstable.  This happens
because the system is trying to evolve from conjunctions at both
planets' pericenters, to conjunctions at the outer planet's pericenter
and the inner planet's apocenter.  Somewhere in between, then, the two
bodies will reach a conjunction at the point where their orbits cross,
bringing about a very close encounter or even a physical collision
\citep[][and Lee \& Peale, private communication]{manhoi_dda_2003}.
However, if a significant inclination between the two orbits has been
attained prior to this transition, as is clearly the case in
Fig. \ref{baseline1}, this will serve as a protection mechanism
against close encounters, allowing the system to stably evolve through
the transition in $\theta_{e_1}$'s libration center.

The simulation is stopped at a little after $7 \times 10^5$ years,
when the inner body comes within 0.1 AU of the star.  
Other runs with randomly varied initial orbital phases, as
well as different intial positions of the planets (though still
starting outside the 2:1 mean-motion resonance) produce a similar
evolution; in each case the the inclination resonance is encountered
after capture into the 2:1 eccentricity resonance, and inclinations
grow rapidly thereafter.  

\section{Dependence on migration rate and initial inclination}

Another example is shown in Fig. \ref{slow_migr}.  In this case, we
reduce the speed at which the disk edge moves inward, $\dot{r}_{\rm
accr}$, by an order of magnitude, which amounts to letting
$\alpha=10^{-4}$. The evolution continues to be very similar to that
shown in Fig. \ref{baseline1}, scaled up by a factor of ten in time.
However, in slowing the migration rate we have also made the migration
more adiabatic; that is, we have reduced the amount of migration
during one libration period of the resonance(s).  Thus, it is to be
expected that the dynamics change by more than a simple rescaling in
time.  One difference apparent from Fig. \ref{slow_migr} is that
libration amplitudes are reduced; the average values of $e$ and $i$,
however, are very close to those in the faster migration.  Another
difference is that the times at which the inclination resonances are
entered and exited do not scale exactly with the slower migration
speed.  

\begin{figure}
\begin{center}
\includegraphics[width=6.0in]{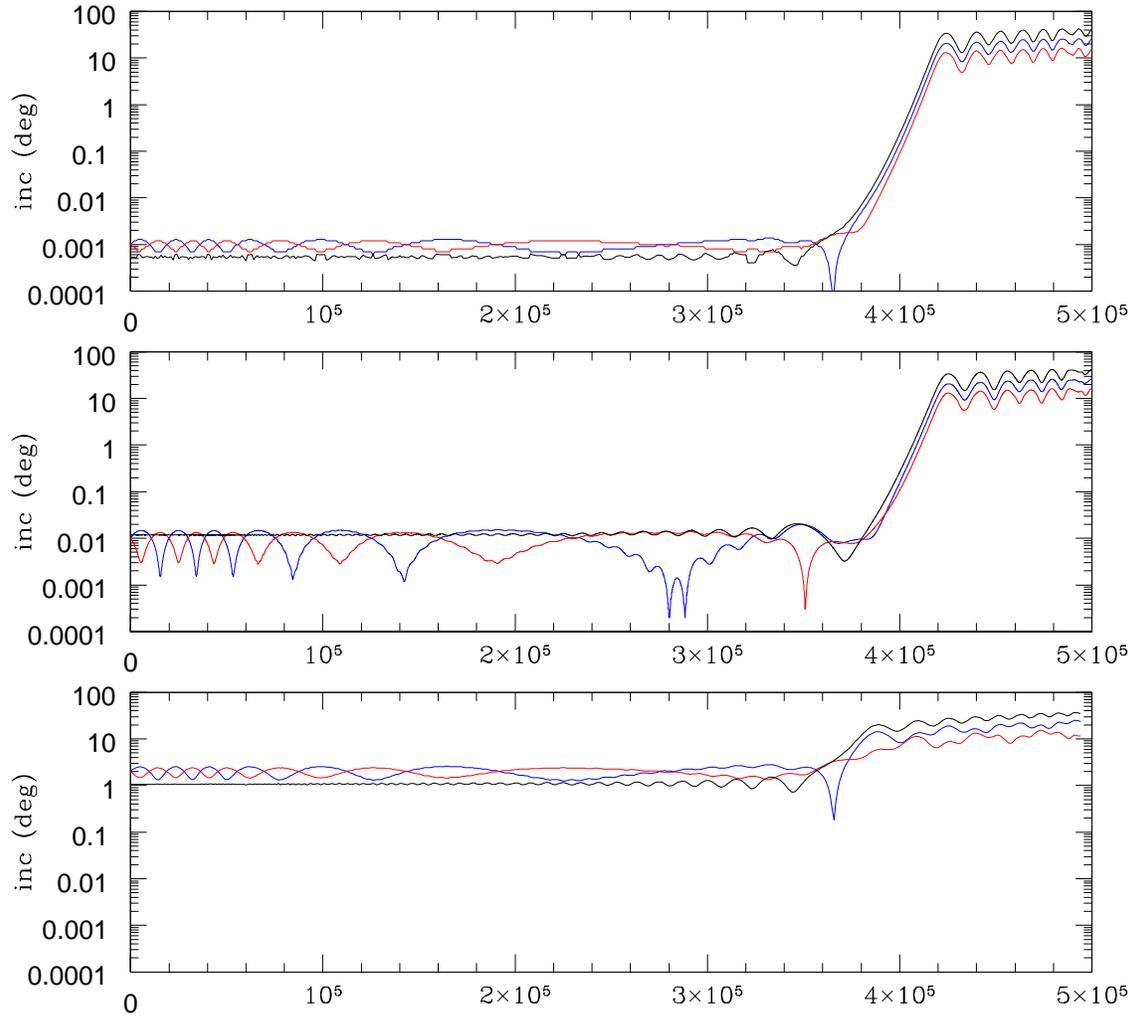}
\caption{Evolution of the planets' inclinations for differing initial
mutual inclinations of $< 0.001\degr$ (top), $0.01\degr$ (middle;
the baseline case), and $1 \degr$ (bottom)} 
\label{different_init_inc}
\end{center}
\end{figure}

We also probe the dependence of the system's evolution on the initial
mutual inclination of the two orbits.  Up to now, we have made this
$\sim 0.01\degr$.  We perform a run where the mutual inclination is
decreased to $\sim 0.0005\degr$, and another where it is increased to $1
\degr$.  The resulting inclination evolution of the two planets is
plotted in Fig. \ref{different_init_inc}, together with the baseline
case of the previous section.  The smaller initial inclination causes
very little change in the time of onset of the inclination excitation,
the timescale over which the inclination is subsequently excited
($\sim 2 \times 10^4$ years), or the value to which the mutual
inclination is subsequently excited ($\sim 30 \degr$).  The larger
initial inclination causes a slightly earlier onset of inclination
growth, by an amount $\Delta t/t \sim 0.05$.  Evidently, therefore,
entry into the inclination resonance is not strongly influenced by the
actual value of the initial mutual inclination, as long as it is
small, $\la 1\degr$.

\section{Varying the mass ratio}
\label{mass ratio} 

\begin{figure}
\begin{center}
\includegraphics[width=6.0in]{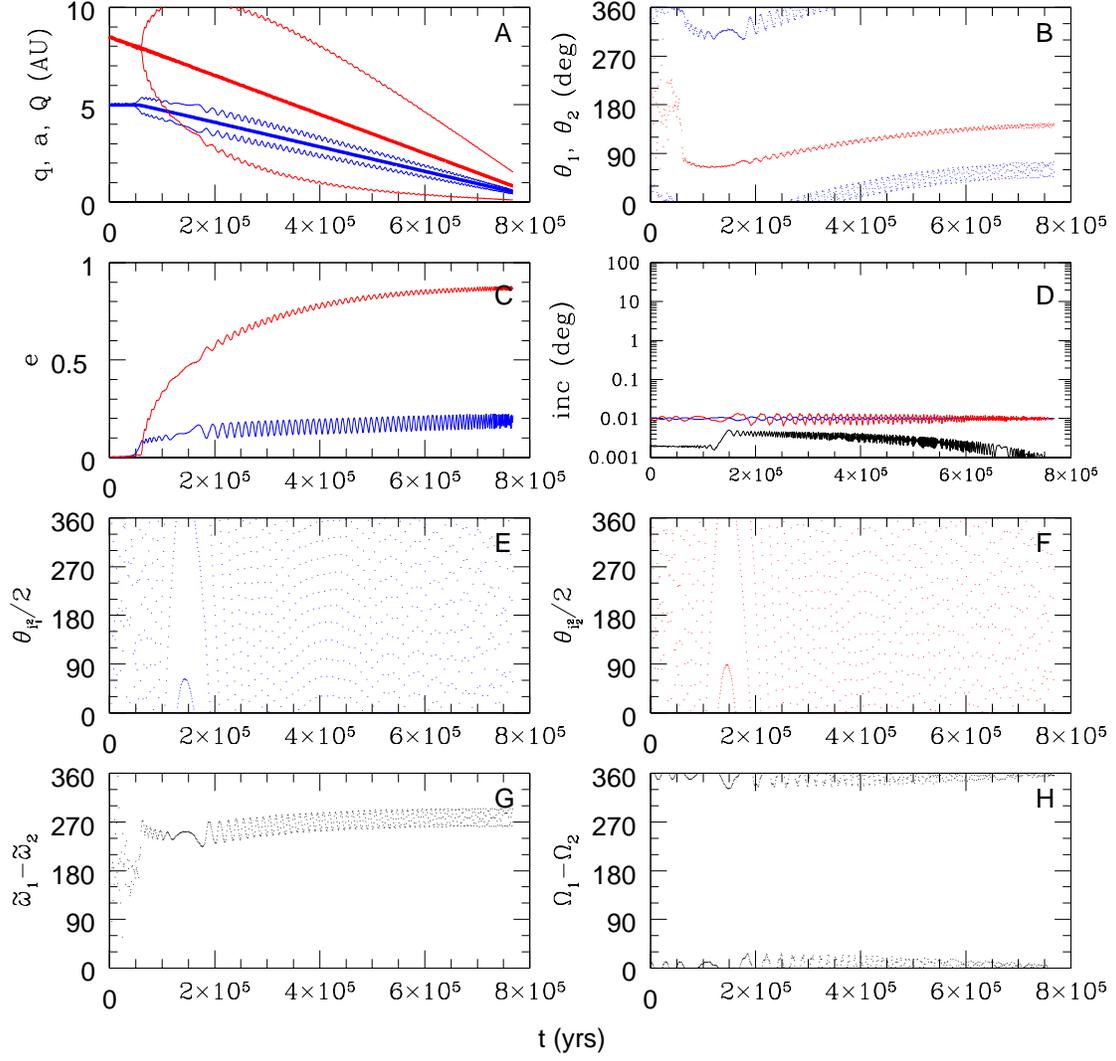}
\caption{Evolution of the planets in the case where the inner is
Jupiter-mass, and the outer is one third that, i.e., Saturn-mass (Section \ref{mass
ratio}). The panels are as in Fig. \ref{baseline1}.}
\label{outersmall}
\end{center}
\end{figure}

We next investigate the effect of making the mass ratio of the two
planets, $M_1/M_2$, differ from unity.  We find that when $M_1/M_2 \la
2$, the inclination resonance is no longer reached.  This is because
onset of the resonance requires the inner body to have a large
eccentricity; however, when there is a large disparity in the masses
of the two planets, the eccentricity of the more massive one will not
grow very much.  Fig. \ref{outersmall} shows an example in which
$M_1/M_2=3$.  Although the outer body's eccentricity has reached 0.85
by the end of the run, that of the inner body does not get above 0.2,
and the inclination resonance is avoided.  We defer a detailed
investigation of the conditions for onset of the inclination resonance
to future work.  For the time being, it is useful to recall that the
simultaneous libration of a body in the 2:1 eccentricity-type
resonance and the 4:2 inclination-type resonance requires that its
longitude of pericenter precession rate, $\dot{\tilde{\omega}}$, be on
average equal to the precession rate of its line of nodes,
$\dot{\Omega}$ \citep[as in a Kozai resonance,
e.g.,][]{1999..Murray..Dermott..book}.  Both of these rates change as
the eccentricity increases; for a body already in the
eccentricity-type resonance, we expect that the inclination-type
resonance sets in when the rates become equal.

\begin{figure}
\begin{center}
\includegraphics[width=6.0in]{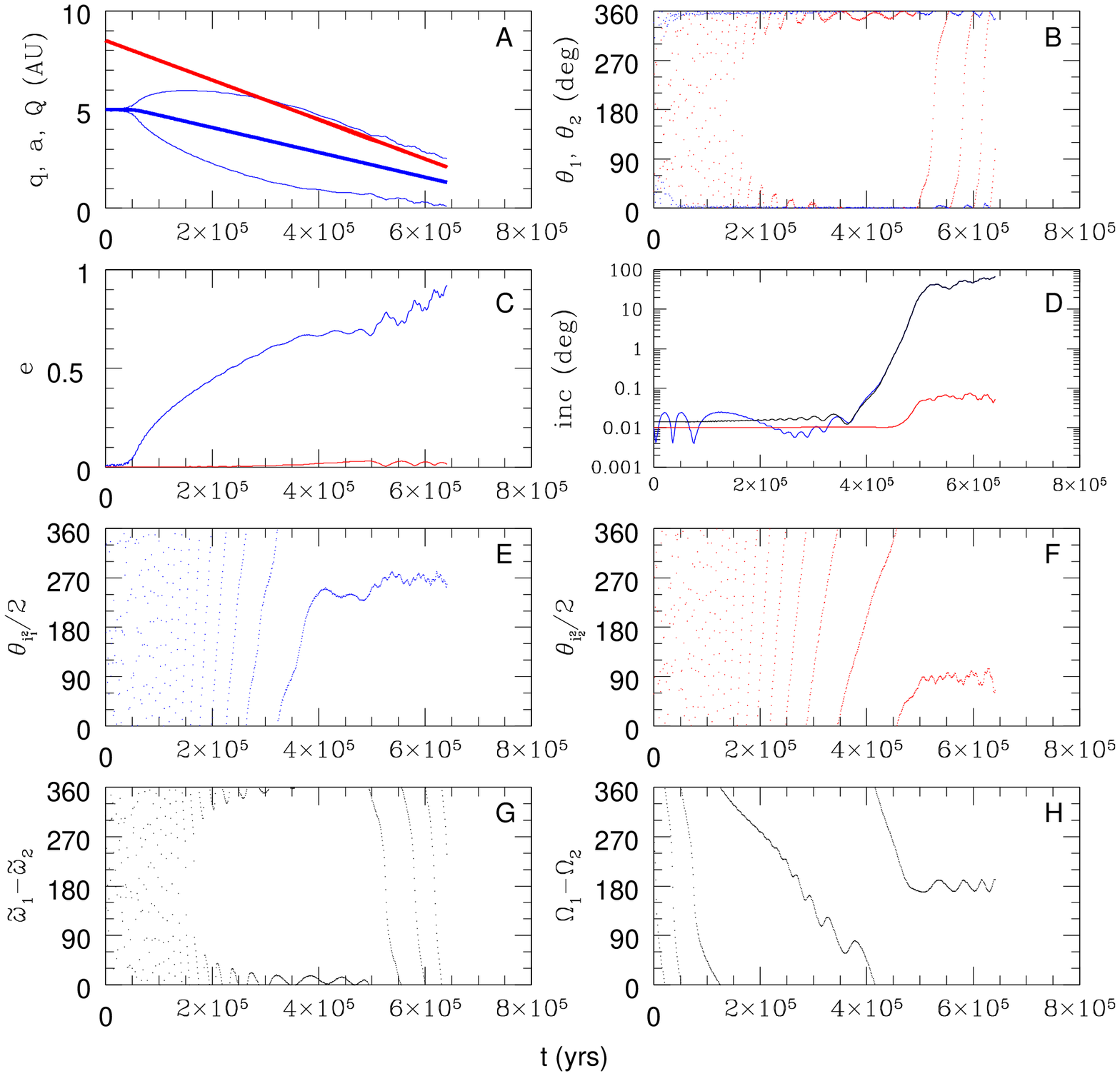}
\caption{Evolution of the planets in the case where the outer planet
is Jupiter-mass, and the inner $\sim 1/300$ Jupiter mass, i.e.,
Earth-mass (Section \ref{mass ratio}). The panels are as in
Fig. \ref{baseline1}.}
\label{innersmall}
\end{center}
\end{figure}

When the inner body is smaller, the inclination resonance is still
entered for much more disparate masses.  Fig. \ref{innersmall} shows
the evolution when the outer planet is Jupiter-mass and the inner is
Earth-mass, making $M_1/M_2 \sim 1/300$.  Though the behaviour is
qualitatively similar to the equal-mass case (Fig. \ref{baseline1}),
there are some obvious differences: As is to be expected from
conservation of momentum, the small planet suffers far more
inclination excitation than the large one.  Also, there is a delay of
$\sim 5 \times 10^4$ years between the onset of libration in
$\theta_{i_1^2}$ and $\theta_{i_2^2}$, and the inclination growth,
which begins with the libration of $\theta_{i_1^2}$, is not as fast as
in the equal-mass case.  

We perform additional simulations in which $M_1/M_2 \sim 10^{-8}$,
making it effectively a test particle (with $M_2$ still being Jupiter
mass, $M_1$ is now comparable to the largest asteroids).
\citet{2001AJ....121.1736Y} find that a test particle in a 2:1
resonance with an inward-migrating massive body, initially inclined
relative to the massive body by of order $1 \degr$, can have its
inclination excited all the way up to $90\degr$ and then be released
from the 2:1 resonance altogether.  When we use a similar initial
inclination, we get capture into inclination resonance and excitation
of the small body's inclination.  We only follow the evolution of the
system until the inner body's pericenter drops below 0.1 AU; at this
point the body is still trapped in both the eccentricity and
inclination resonance, and its inclination, which is still growing,
has reached $\sim 70\degr$.  It is now only $\theta_{e_1}$ and
$\theta_{i_1^2}$, not $\theta_{e_2}$ and $\theta_{i_2^2}$, which
librate.  With an initial mutual inclination of $0.01\degr$, however,
it is possible for the small body to quickly pass through the
inclination resonance, ``librating'' for less than a full cycle before
leaving it again, while having its inclination excited to only $\sim 1
\degr$.  Though \citet{2001AJ....121.1736Y} do not check for libration
in inclination-type resonances, the similarity in evolution of
eccentricity and inclination between our simulations and their Fig. 5
leads us to believe that the behaviour they see is the
$M_1/M_2\rightarrow 0$ limit of the mechanism we explore here.  In any
case, we do not investigate the behaviour at such extreme mass ratios
in more detail in the present work; if we are to consider pairs of
giant planets, the possible masses range from $10^1$ to $10^3$
M$_{\oplus}$, so the smallest possible ratio is of order $M_1/M_2 \sim
10^{-2}$, as in the case pictured in Fig. \ref{innersmall}.

\section{Damping of eccentricities}
\label{damping}
Next, we consider the possibility that planet-disk interactions may
damp planetary eccentricities.  Following \cite{2002ApJ...567..596L},
we make the relative eccentricity damping rate proportional to the relative
{\it applied} semimajor axis decay rate for the outer planet:
\begin{equation}
\frac{\dot{e}_2}{e_2} = S \frac{\dot{a}_2^{\rm appl}}{a_2}.
\label{damping1} 
\end{equation}
The applied decay rate, $\dot{a}_2^{\rm appl}$, is the rate at which
the outer planet by itself would migrate under the action of the
applied torque.  However, because the outer planet has to push along
the resonantly-locked inner planet, its actual migration rate,
$\dot{a}_2$, is less than $\dot{a}_2^{\rm appl}$.  Recall that
torque is applied in such a way as to keep $\dot{a}_2$ (nearly)
constant (Eq. \ref{gap prescription}).

\begin{figure}
\begin{center}
\includegraphics[width=6.0in]{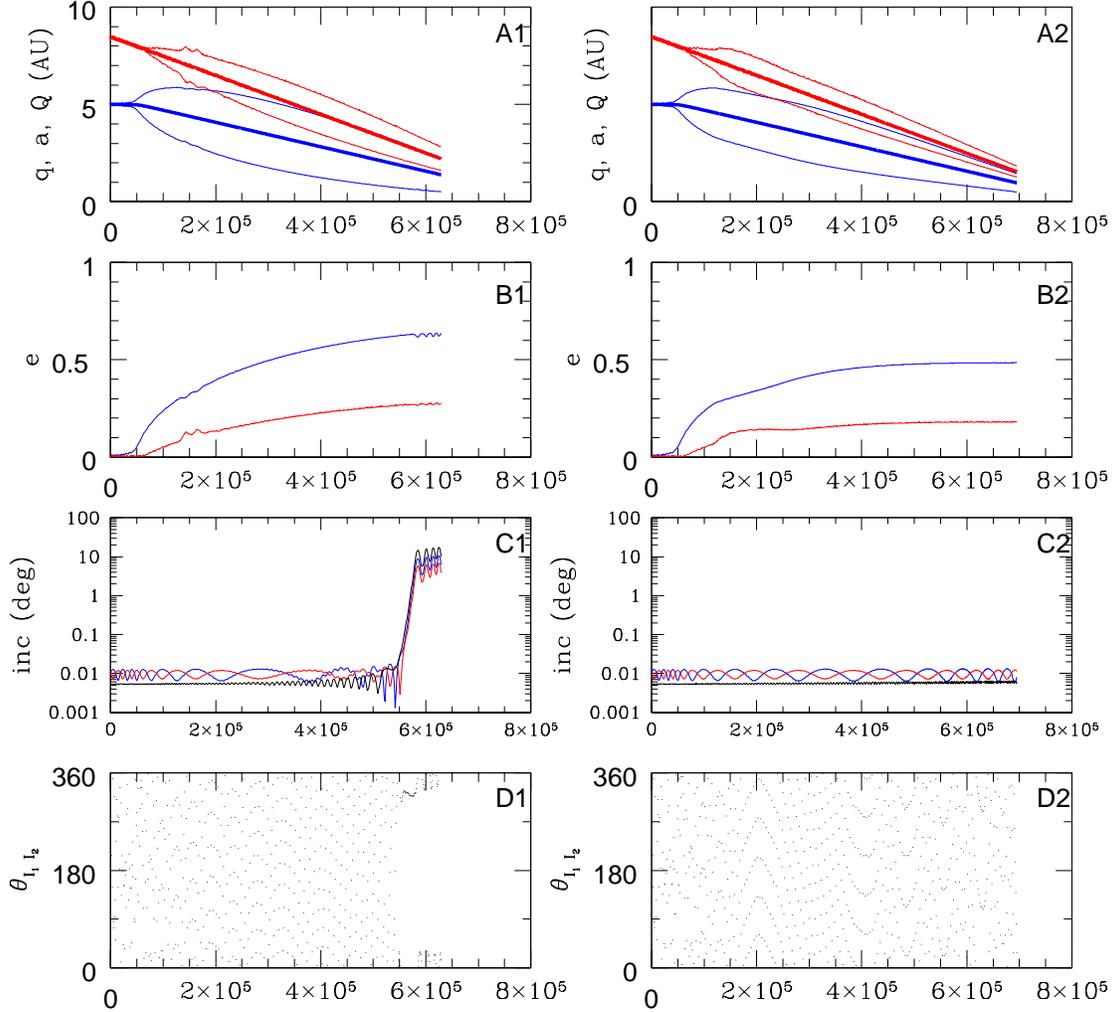}
\caption{Evolution of the two-planet system when the eccentricity of
the outer is damped.  The left set of panels show the case where the
relative damping rate, $\dot{e}_2/e_2$, is twice the applied relative
semimajor axis decay rate, $\dot{a}_2^{\rm appl}/a_2$; the right set
show the case where the relative damping rate is $5\times$
$\dot{a}_2^{\rm appl}/a_2$.  The individual panels show: A.  Semimajor
axis, periastron and apastron distance; B.  Eccentricities;
C. Inclinations; D. The mixed inclination-type resonance angle
$\theta_{i_1i_2}$}
\label{damping_paper}
\end{center}
\end{figure}

This eccentricity damping rate is applied to the outer planet only;
the inner is, again, assumed to not be interacting with the disk
directly.  We perform simulations with $S=2$ and $S=5$.  Results are
shown in Fig. \ref{damping_paper}.  In the $S=2$ case, the inclination
resonance is entered later due to the slower eccentricity growth, and
the maximum mutual inclination reached is less than $20 \degr$.  When
$S=5$, the planets' eccentricities reach equilibrium values before the
inclination resonance is attained, and the system thus remains
coplanar.

One effect we have not modeled is direct damping of the planets'
inclinations by the disk.   \cite{2001ApJ...560..997L} find that such damping
should occur when $\alpha \ga 10^{-3}$, and proceed on a timescale of
order $10^5$ years.  Thus, our baseline runs, with $\alpha = 10^{-3}$,
are at the limit of where damping is expected to occur.  Also, the
inclination damping timescale is long compared to the timescale of
inclination growth when the inclination resonances are first
encountered, which is only a few $\times\, 10^4$ years.  Thus, damping
of inclinations by the gas disk likely has little effect on
inclinations excited by resonances among planets. 

\section{The disk potential}
\label {disk potential} Thus far we have only included the effect
of the gas disk in the form of a forced migration experienced by the
outer planet. However, the disk may influence the dynamics of the
planets in other important ways.  Specifically, the gravitational
force from a disk will modify the planets' nodal and apsidal
precession rates, $\dot{\Omega}$ and $\dot{\tilde{\omega}}$.  This can
change the locations and relative spacings of the mean-motion
resonances, and so might change the nature of the system's
resonant evolution.  
Therefore, it is important to incorporate the
potential of the disk into our analysis.  We assume that the
``disk'' is a thin annulus of surface density $\Sigma(r)$ located in
the $z=0$ plane, with inner
radius $r_1$ and outer radius $r_2$, where $r$ is the
stellocentric distance. The potential due to
the disk, as felt at a radius $R$ and height $z$, is
\begin{equation}
\Phi(R,z) = -\int^{r_2}_{r_1}{\int^{2 \pi}_0 {\frac{G \Sigma(r) r d\phi
dr}{\sqrt{R^2 + r^2 + z^2 -2 r R \cos\phi}}}},
\label{diskpotential_1}
\end{equation}
and the corresponding force is $F_R(R,z) = -\partial \Phi/\partial R$,
$F_z(R,z)=-\partial \Phi/\partial z$.  To obtain the force due to an
outer disk on a body, we can expand the numerator in powers of $R/r$
and $z/r$,
then perform the integration.  Assuming a surface density of
the form
\begin{equation}
\Sigma(r) = \Sigma_0 r^{-3/2},
\label{surface_density}
\end{equation}
i.e., having the density profile of the standard
\cite{1981PThPh..70...35H} nebula model, we obtain

\begin{equation}
F_R(R,z) = \left . \frac{G \Sigma_0 \pi}{r^{3/2}} \left
(-\frac{2}{5}(R/r) + (R/r)^3 - \frac{2}{3}(R/r)(z/r)^2+... \right )
\right \vert^{r=r_2}_{r=r_1}
\label{fr_outer}
\end{equation}
and
\begin{equation}
F_z(R,z) = \left . \frac{G \Sigma_0 \pi}{r^{3/2}} \left
(\frac{4}{5}(z/r) + (R/r)^2 (z/r) - \frac{2}{3}(z/r)^3+... \right )
\right \vert^{r=r_2}_{r=r_1}.
\label{fz_outer}
\end{equation}

We performed a simulation with a disk surface density of 5 times
$\Sigma_{min}$, where
\begin{equation}
\Sigma_{min} = 1.7 \times 10^3 (r/\mbox{1 AU})^{-3/2} \mbox{ g/cm}^2
\label{minmass}
\end{equation}
is the Hayashi minimum-mass surface density.
\begin{figure}
\begin{center}
\includegraphics[width=6.0in]{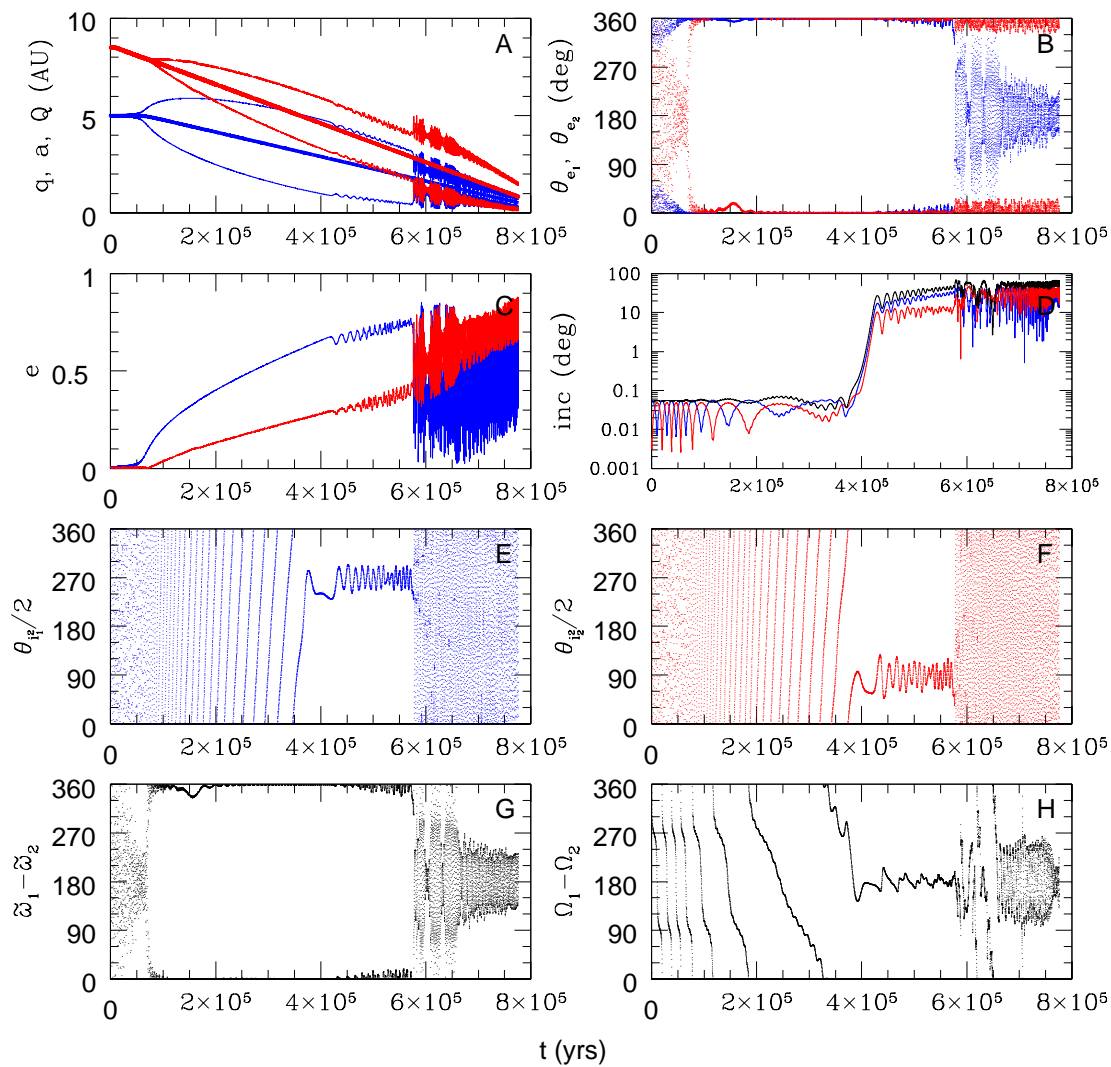}
\caption{Evolution of the two-planet system with the addition of
the potential of a $5 \times$ minimum-mass outer disk.  Panels are
as in Fig. \ref{baseline1}. } \label{5minmass1}
\end{center}
\end{figure}
The inner radius of the disk is made to be 20 Hill radii of the outer
planet beyond $r_{\rm disk}$ at all times.  The Hill (or Roche)
radius of a planet of mass $M$ orbiting at a distance $r$ from a primary
of mass $M_*$ is $r_H = (M/3 M_*)^{1/3}r$.  The actual width of a
planet-induced gap is likely significantly smaller, between 3 and 4
Hill radii \citep*[e.g.,][]{2000ApJ...540.1091B}.  However, since the
migrating planets tend to acquire large eccentricities, we endeavour
to keep the disk edge far enough away that it lies well beyond the
planets' apocenters at all times, thus keeping the expansion of the
potential valid.  The response of a disk to an eccentric gap-opening
planet is a complicated problem which has thus far not been
investigated \citep[e.g.,][]{2000prpl.conf.1111L}, so to avoid having to make
arbitrary assumptions, we restrict ourselves to modeling the
gravitational interaction of the planets with the more distant part of
the disk.

The evolution of the system is shown in Fig. \ref{5minmass1}.
Comparing to Fig. \ref{baseline1}, one can see that the evolution
is quite similar to the diskless case.  Despite the additional
radially-varying apsidal precession induced by the disk potential,
the two planets again become apsidally aligned upon encountering
the 2:1 resonance.  Also, despite the change in the vertical
frequency, the inclination resonance sets in at nearly the same
time as in Fig. \ref{baseline1}.  One noticeable difference is
that the angle between the planets' lines of nodes,
$\Omega_1-\Omega_2$, has a larger amplitude of libration about
$180\degr$, and in fact circulates briefly several times. This is
possible because, with the addition of the disk potential, the
two-planet system no longer conserves the component of its angular
momentum which lies in the disk plane.

\section{Additional planets}
\label{addl planets}
Finally, we investigate the effect that additional smaller planets
have on migration in resonance.  We add four bodies, each 15
M$_{\oplus}$ in mass (i.e., Uranian or ``ice giant'' size), interior
to the inner of the two Jupiter-mass bodies.  The bodies are separated
by between 10 and 15 mutual Hill radii, so as to form a relatively
stable system.  The innermost initially has a semimajor axis around 1
AU, while the outermost is at 3 AU, just short of the interior 2:1
resonance of the Jovian body at 5 AU. We do not argue that it is
likely for a planetary system's inner region to contain so many large
bodies at any time in its evolution.  In-situ formation alone is, at
any rate, unlikely to produce such massive bodies in this region
\citep*[e.g.,][]{1987pggp.rept..100L,2003Icarus...tdl...oligarchy}.  Rather, our intent is to
put into the simulation about as much mass in sub-gas giant bodies as
will stably fit in the region.  In this way, we obtain an idea of the
upper limit (apart from stochastic close approaches) on the
perturbations that the inner part of the planetary system could exert
on the resonant migration of the two gas giants.

\begin{figure}
\begin{center}
\includegraphics[width=6.0in]{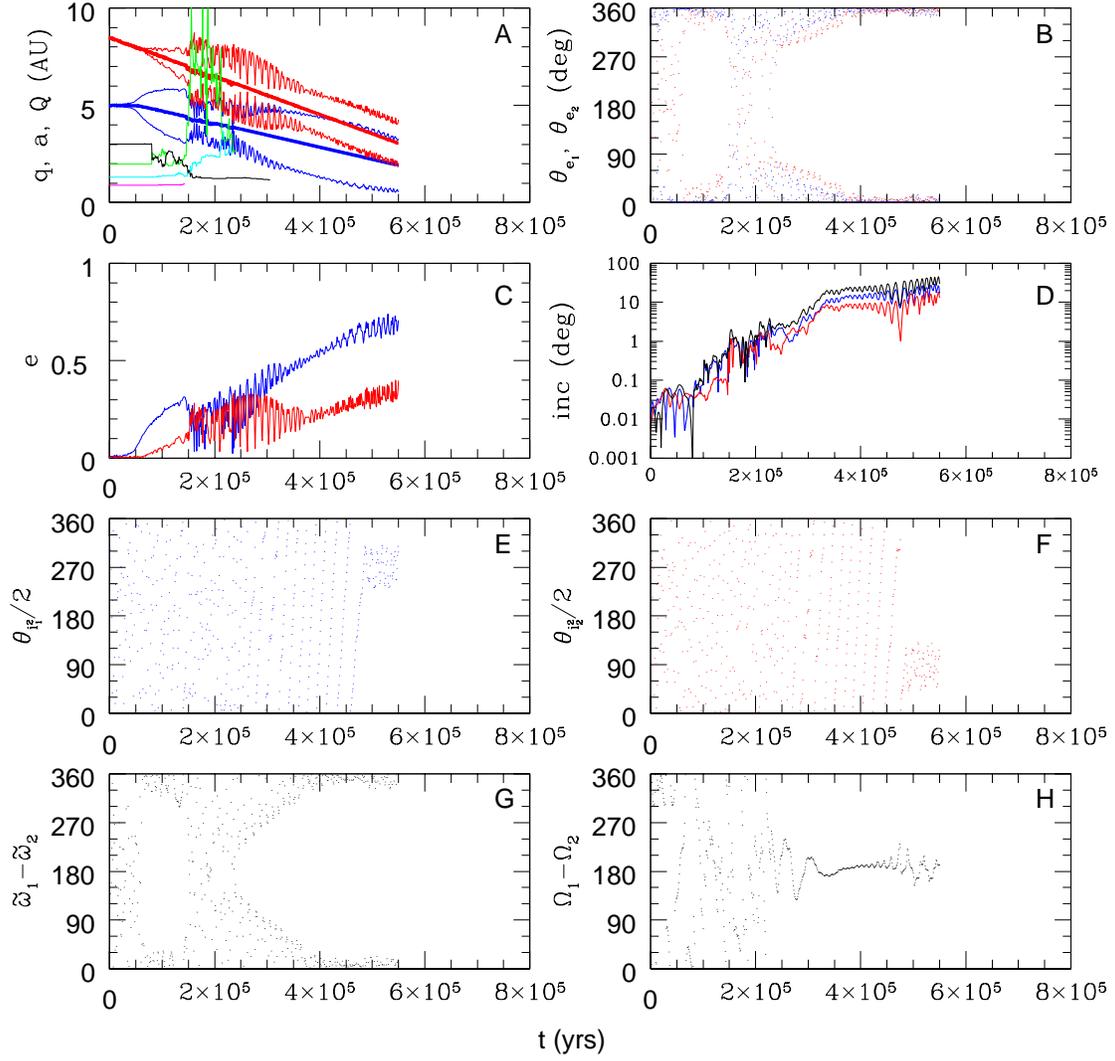}
\caption{An example of the evolution of the system with four 15
M$_{\oplus}$ (Uranian mass) planets added interior to 3 AU. Panels are
as in Fig. \ref{baseline1}, and additionally the semimajor axes of the
Uranian planets are shown (in black, green, cyan and magenta) in Panel
A.  The last of them is eliminated at about $3 \times 10^5$ years.
}\label{inner_uranians}
\end{center}
\end{figure}

Fig. \ref{inner_uranians} shows the resulting orbital evolution,
including semimajor axes of the Uranian-mass bodies, in one of a
set of eight simulations performed.  The last of these extra
bodies is eliminated at 3$\times 10^5$ years.  All are removed by
crossing the inner simulation boundary, which in these runs is at
0.5 AU. Comparing to Fig. \ref{baseline1}, there are some key
differences between the orbital evolution of the two gas giants
here and in the baseline simulation.  Here, the onset of the
inclination resonance, shown by the transition of $\theta_{i_1^2}$
and $\theta_{I^2_2}$ from circulation to libration, takes place
about $2 \times 10^5$ years later than in the baseline simulation.
However, the gas giants already acquire significant inclinations
before this happens.  The growth of the inclinations begins at
about the same time as $\theta_{\Omega}$, the difference between
the lines of nodes, begins to librate about $180\degr$.  This
means that, although the two planets are not in the pure
inclination-type resonances, they are (since $\theta_{e_1}$ and
$\theta_{e_2}$ are librating at the same time) in the mixed
eccentricity-inclination resonances involving the libration of
$\theta_{e_1i_1i_2}$ and $\theta_{e_2i_1i_2}$.

The pure inclination-type resonances do not set in until after the
last Uranian planet is eliminated, and final inclinations are
comparable to those reached at the same time in the baseline run.
The transition from libration about $0\degr$ to libration about
$180\degr$ of one of the eccentricity-type resonances seen in Fig.
\ref{baseline1} does not occur here, simply because the inner
boundary in this run is at 0.5 AU instead of 0.1 AU (to allow a
longer timestep and thus speed the integration), and therefore the
inner Jovian planet is eliminated before its libration center can
switch.  

\begin{figure}
\begin{center}
\includegraphics[width=6.0in]{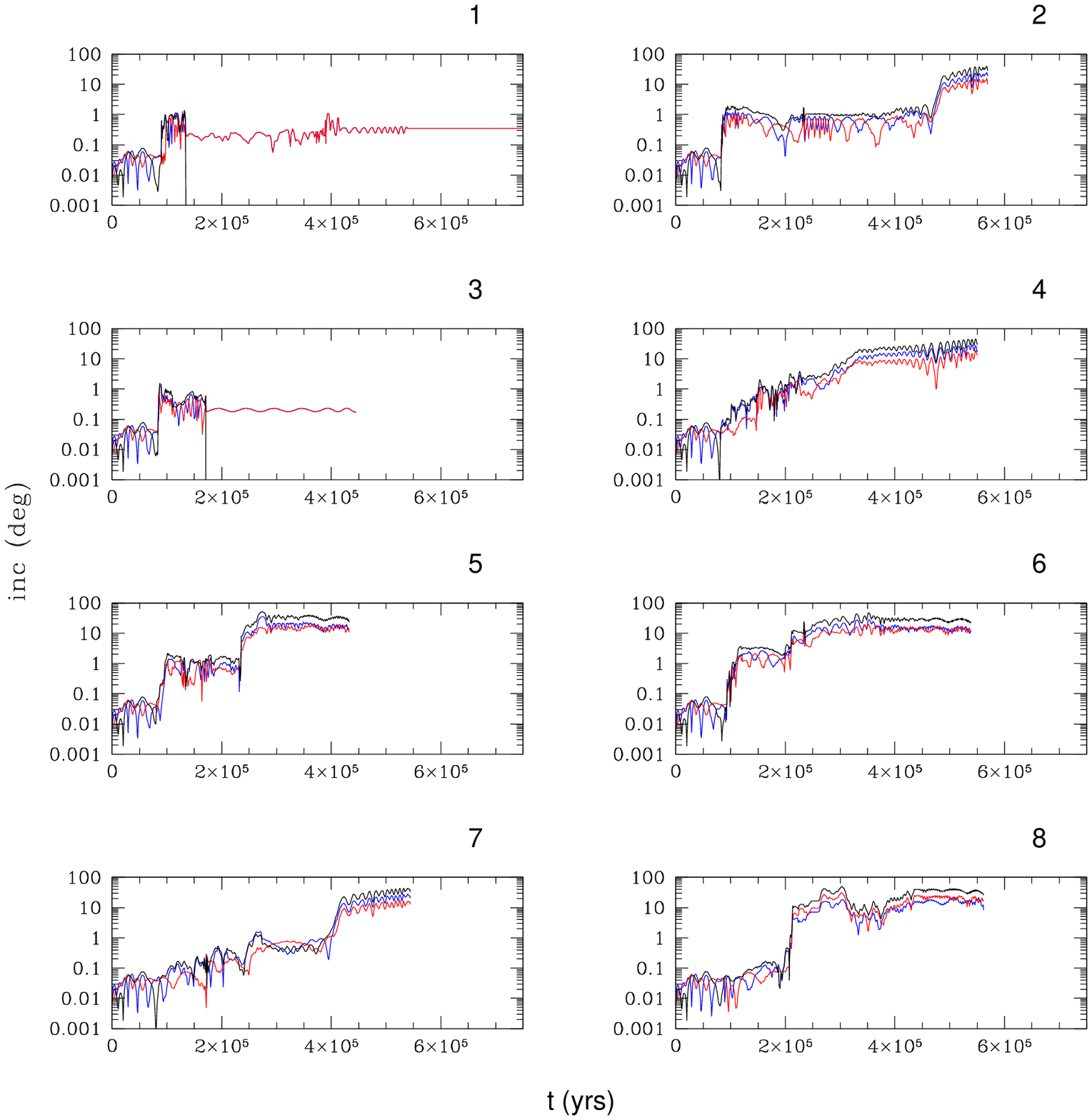}
\caption{The full set of eight runs performed with four inner
Uranian planets, showing the evolution of the two Jupiter-mass
planets' inclinations in each case.  In runs 1 and 3, the two
Jovian planets merge before the inclination resonances are
encountered.} \label{all_inc_plot}
\end{center}
\end{figure}

With this many bodies, the system becomes strongly stochastic.  Of the
eight runs performed, six evolve in a manner qualitatively similar to
the one shown in Fig. \ref{inner_uranians}: The gas giants undergo
resonant capture and evolve inward together, pushing the interior
planets ahead of themselves.  In all of these cases, significant
resonant excitation of the giant planets' inclinations occurs, though
in some cases a 3:1 rather than a 2:1 commensurability is involved. In
the two remaining runs, the gas giants undergo a physical collision.
This happens because close encounters with the smaller planets can
exert a strong enough perturbation on the large planets to temporarily
knock them out of resonance with each other.  The large planets'
orbits already cross by this time, so while their resonant lock is
removed, they are free to undergo close encounters with
each other.

Fig. \ref{all_inc_plot} shows the evolution of the inclinations in
each of the eight runs.  Runs 1 and 3 are the ones where the gas
giants collide, in both cases before significant inclinations are
excited.  In all of the other six runs, relative inclinations in
excess of $30 \degr$ arise.

\section{Discussion}
\label{discussion}
We have shown that two planets migrating in resonance can undergo
significant excitation of their inclinations under a wide range of
conditions.  The excitation mechanism is libration in the lowest order
(order $i^2$) inclination-type mean-motion resonances.  Even though
convergent migration initially leads to capture into a first-order
eccentricity-type resonance, at sufficiently large eccentricity, the
two bodies can librate simultaneously in both types.  Previous work
has also suggested the possiblity of high inclinations arising during
the formation of planetary systems
\citep*{1998AJ....116.1998L,2001AJ....121.1736Y}.  In the former case,
the inclination excitation mechanism considered was stochastic close
encounters among planets.  However, since these typically eventually
lead to ejections and mergers, one would expect few long-lived
non-coplanar systems of closely-spaced planets to be produced in this
way.  In the latter case, the authors showed that massless test
particles resonantly trapped by an inward-migrating planet can
eventually be released from resonance on polar and even retrograde
orbits; this suggests the possibility of non-coplanarity between giant
planets and much smaller bodies.  We think it is likely that the
effect seen by the authors is actually the $M_1/M_2 \rightarrow 0$
limit of the mechanism describe herein (see Section \ref{mass ratio}).

Resonant inclination excitation is quite robust against external
perturbations.  Adding the potential of a fairly massive (5 times
minimum mass) outer gas disk into the simulation does not prevent the
onset of the inclination resonance.  Neither, in most cases, does the
addition of a massive inner planetary system. However, it appears that
the inclination resonances cannot be encountered until the
eccentricity of the inner planet gets quite high, $e_1\ga 0.6$. Thus,
damping of the eccentricity can prevent excitation of the
inclinations.  Given equal-mass bodies in a 2:1 resonance, we find the
threshold relative damping rate for the outer body, $\dot{e}_2/e_2$,
is between $2$ and $5$ times the relative applied migration rate,
$\dot{a}_2^{\rm appl}/a_2$ (the inner planet is assumed not to
interact with the disk).  Also, if the mass ratio $M_1/M_2$ is larger
than about 2, the inner body undergoes very little eccentricity
excitation and the inclination resonances are avoided.

There is no obvious reason why having two adjacent companions with
$M_1/M_2 < 2$ should be an uncommon feature of planetary systems.  In
fact, the majority of discovered multiplanet systems, and both of the
systems in 2:1 resonance (see below), fulfill this condition at least
with their projected masses, $M \sin i$.  Our own Solar System, it
should be pointed out, does not ($M_{\rm Jupiter} \approx 3 M_{\rm
Saturn}$).  Less clear, though, is the issue of the disk's effect on
planet eccentricities.  There is not even a consensus on whether the
disk-induced eccentricity rate of change,
$\dot{e}_{\rm disk}$, is positive or negative.
\citet*{2002...goldreich...sari} analytically show that the former
ought to be the case.  However, numerical simulations have
tended to support the latter; for example,
\citet{2000MNRAS.318...18N} perform simulations of gap-opening
planets with three different hydrodynamic codes and find
$\dot{e}_{\rm disk}$ to be negative in each case. Also, in the simulations
of \citet*{2001A&A...366..263P}, disk interactions do not cause
eccentricity growth until the planetary mass exceeds $\sim 20$ M$_J$.
The problem is that the competing effects of eccentricity damping at
corotation resonances and eccentricity pumping at Lindblad resonances
are very nearly of equal strength.  Whether one or the other dominates
depends sensitively on the details of the disk's surface density
profile in the vicinity of the planet, which is a difficult issue to
handle either numerically or analytically.  In any case, an
eccentricity damping timescale of the same order as the orbital decay
timescale or longer, which we find is required to permit capture into
inclination resonances, is entirely plausible.  A related question is
whether the disk might impose some maximum eccentricity on embedded
planets.  As previously mentioned, the standoff distance between a
disk and a gap-opening planet's orbit is between 3 and 4 Hill radii in
numerical simulations \citep[e.g.,][]{2000ApJ...540.1091B}.  Assuming
this distance is independent of planetary eccentricity, a Jupiter-mass
planet will run into the inner disk edge at apocenter when its
eccentricity is $\sim 0.2-0.3$, which may lead to increased damping.
In all of the simulations performed here (except for the one shown in
Fig. \ref{outersmall}, but there the inclination resonance was
avoided anyway), the outer planet's eccentricity is $\sim 0.3$ or less at the
time the inclination resonance is entered.  Thus, if such a gap-edge
cutoff exists, a pair of resonant planets still has a good chance of
entering the inclination resonance before suffering significant
eccentricity damping.

We find that once capture into the inclination resonances sets in, the
inclinations of the planets can rapidly grow to $\sim 15\degr$
relative to their plane of origin.  Since the lines of nodes of the
two planets' orbits are anti-aligned after entering the inclination
resonance, the relative inclination is the sum of each orbit's
inclination, and grows to $\sim 30 \degr$.  Later, when the system
undergoes large coupled oscillations in eccentricity and inclination,
mutual inclinations can get as high as $\sim 60 \degr$
(Fig. \ref{baseline1}).  Thus, a planetary system which enters
inclination resonance during its evolution may end up in a highly
non-coplanar configuration.  This is an important possiblity to
consider in dynamical investigations of exoplanetary systems; in fact,
even the inclination relative to our line of sight, and thus the $\sin i$
scale factor for the planetary masses, may differ significantly among individual members of
multi-planet systems.

Among the discovered exoplanetary systems, the best candidates for
having mutually-inclined orbits due to the mechanism described here
are those in which the companions are (thought to be) in a mean-motion
resonance.  In the case of the GJ 876 system,
\citet*{2001ApJ...558..392R} perform dynamical fits allowing for the
possiblity of mutual inclinations.  Most of the systems they find
which both provide good fits to the radial velocity data and long-term
stability have mutual inclinations of $12 \degr$ or less.  They also
find one (not necessarily the only) stable, locally best-fit (in the
sense of a local $\chi^2$ minimum) region at a very high mutual inclination of
$77\degr$. Leaving aside this extreme case, since none of the runs performed
here result in such a high mutual inclination for any significant
length of time, one might thus conclude that GJ 876 {\it could} have
undergone resonant inclination excitation, provided some
mechanism---perhaps eccentricity damping (see
Fig. \ref{damping_paper}), or direct damping of inclinations by the
disk, which we have not considered---kept inclinations from getting
very high.  Also, the outer companion's $M \sin i$ is about 3.4 times
that of the inner, so if the outer planet is oriented edge-on to our
line of sight (see below), the condition $M_1/M_2 < 2$ is fulfilled
unless the inclination between the two orbits is $\ga 80\degr$.
However, the system's relatively low eccentricities, $\la 0.3$ for
the inner planet, and $\la 0.1$ for the outer
\citep{2001ApJ...556..296M,2001ApJ...551L.109L}, argue against
resonant inclination excitation.  Unless the eccentricities were
significantly higher than this in the past, it is unlikely that the
inclination resonance would ever have been reached.

The two companions of HD 82943 are also thought to be in a 2:1
resonance, and these have inferred eccentricities of 0.41 (outer) and
0.54 (inner) (as announced at an ESO press release, April 4, 2001; see
http://obswww.unige.ch/~udry/planet/hd82943syst.html).  Also, the
outer companion's $M \sin i$ is more than twice that of the inner.
This pair of planets may therefore provide a more likely setting for
resonant inclination excitation.

The actual mutual inclinations in multiple-planet systems should
eventually be revealed by astrometric measurements.  In fact, the
first and thus far only astrometrically-measured orbit is that of the
outer companion of GJ 876 \citep{2002ApJ...581L.115B}, showing its
orbit to be oriented nearly edge-on to the line of sight, and thus
determining its mass to be about $1.9$ M$_{J}$.  The smaller semimajor
axis and the (very likely) smaller mass of the inner companion, GJ
876c, will make it more difficult to detect astrometrically; if
coplanar with GJ 876b, it should produce a perturbation in the star's
position around 0.2 that produced by the outer, or about 0.05
milliarcseconds.  Such astrometric precision is attainable in the
forseeable future; certainly, it lies well within the planned
microarcsecond precision of the Space Interferometry Mission (SIM).

A mechanism for tilting planetary orbits relative to the orientation
of the original protoplanetary disk is particularly interesting in the
context of the $\beta$ Pictoris dust disk
\citep*{1995AJ....110..794K}. The disk, which is oriented nearly
edge-on to the line of sight, displays a warp out to a distance of
about 50 AU; the disk inside this radius is inclined about $3 \degr$
relative to the outer disk. \citet{1997MNRAS.292..896M} modeled the
warp as arising from the influence of a single unseen planet, with a
mass between $10^{-2}$ and $10^{-5}$ that of the central star, with a
semimajor axis between 1 and 20 AU, and inclined 3 to $5\degr$ to the
midplane of the outer disk.  As long as the inclination resonance is
actually reached, the simulations we perform here indicate that such
inclinations could easily arise.  Additional structure revealed in
recent observations of $\beta$ Pic seems even more suggestive:
\citet{2003...wahhaj} conclude from their observations that the disk
in fact contains multiple warps interior to 100 AU, being made up of
four inclined rings which they denote A, B, C and D, centered at
roughly 14, 28, 52 and 82 AU respectively. They estimate that these
these rings are inclined at about $-32\degr$, $+25\degr$, $-2\degr$
and $+2\degr$ respectively, relative to the outer disk. They point out
that the ring radii suggest low-order mean-motion commensurabilities:
3:1 between A and B, and 2:1 between C and D.  There is also the
possiblility of a higher-order commensurability, 7:3, between B and C.
\citet{2003...wahhaj} suggest that these rings are associated with
planets.  Their observations appear to fit well with a scenario like
the one we explore here, in which planets acquire large inclinations
relative to the original plane of the protoplanetary disk by migrating
in resonance with each other.  Particularly intriguing are the
alternating positive-negative inclinations of adjacent rings, mimicing
the nodally anti-aligned configurations that the planets take on in
our simulations once they acquire significant inclinations. 

\begin{figure}
\begin{center}
\includegraphics[width=6.0in]{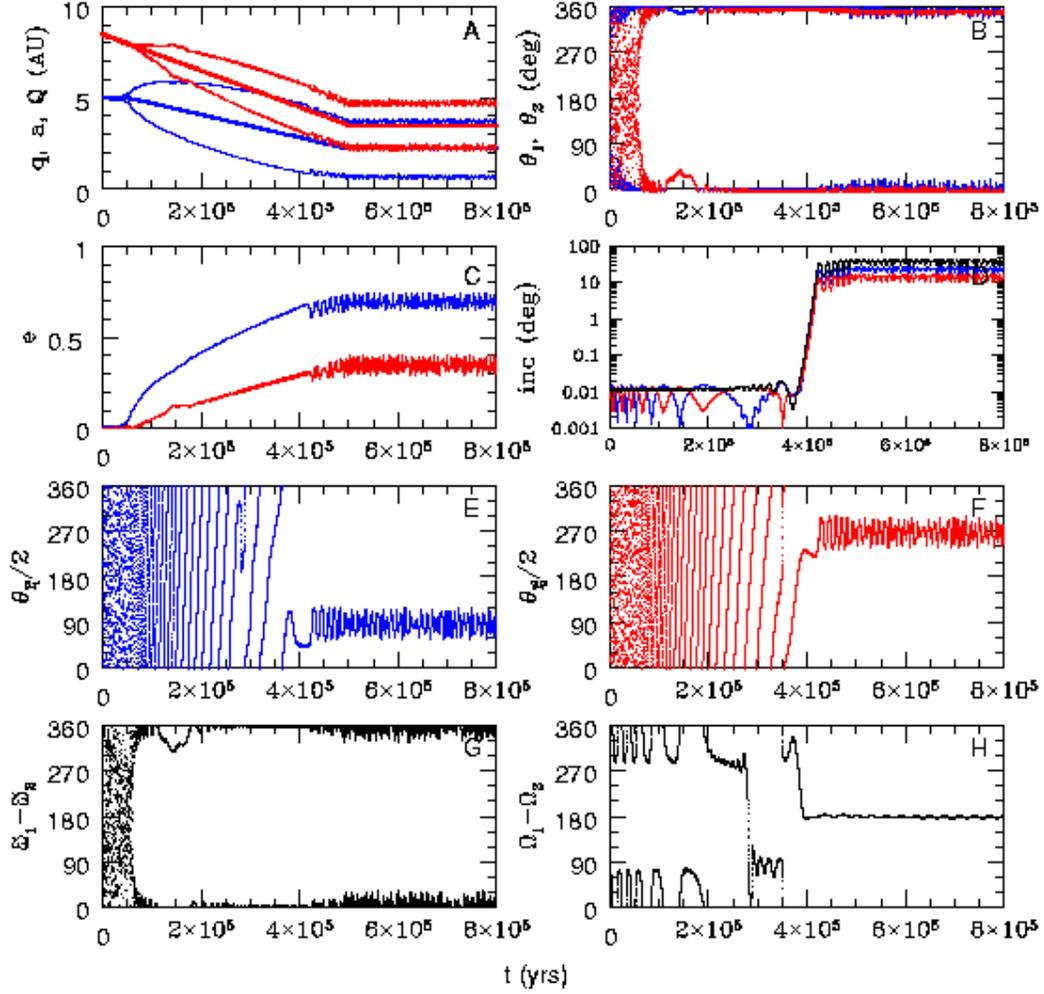}
\caption{Another version of the baseline run (cf. Fig. \ref{baseline1}), this time with migration
turned off at $5\times 10^5$ years. } \label{frozen_migr}
\end{center}
\end{figure}

Thus far, we have not considered the ultimate fate of our
inward-migrating planets.  The simulations are simply stopped once
either planet crosses the inner simulation boundary (0.5 or 0.1
AU). However, the configuration we are ultimately left with is
sensitively dependent on how the migration ``end game'' actually plays
out.  If migration keeps going down to very near the star, one or both
planets will be lost.  Alternatively, it is possible that the gas disk
could dissipate sometime after resonant capture but before the planets
are driven into the star, though this would require a coincidence in
timing.  In the absence of other effects that outlive the gas, such as
interaction with a massive debris disk, interaction with other planets,
and for bodies with very small periastron distances, tidal
interaction with the primary, the mean inclinations (and
eccentricities) should remain at their values at the time migration
ceases.  Fig.  \ref{frozen_migr} shows an example in which we
(abruptly) turn off the induced migration of the outer planet at $6
\times 10^5$ years, shortly after the system has entered the
inclination resonances.

The longer the migration timescale (i.e., the lower the disk
viscosity), the more plausible such a coincidence becomes. Another
possibility to consider is that migration may under some circumstances
be outward, for instance if a resonant pair of planets does not form a
clean gap and mass flows inward across the gap, as demonstrated by
\citet*{2001MNRAS.320L..55M}.  In this case, planets could keep
migrating indefinitely without meeting a catastrophic end, though they
would eventually move beyond the stellocentric radius accessible to
radial velocity searches.

Finally, there is the intriguing possibility that inclination
excitation could actually serve as a ``parking mechanism'' which puts
an end to migration.  One should keep in mind that we have used a
simple, ad-hoc migration prescription in our simulations, nothing more
than a constant, negative $\dot{a}$ which is enforced for the outer
planet. The interaction of a significantly eccentric and inclined
giant planet with a disk is in fact highly uncertain.  One may
speculate that a highly-inclined planet might be coupled less strongly
to the surrounding disk.  At the very least, by the time the planet's
inclination exceeds that of the disk ($\sim 4\degr$ from midplane to
scale height at 5 AU, decreasing inward for a flared disk), it may no
longer be able to maintain a clean gap or edge.  Such questions must
ultimately be addressed through detailed hydrodynamic simulations,
which as yet are restricted to the case of planets on circular, disk-coplanar orbits.

\acknowledgements 

We thank Man Hoi Lee, Stan Peale, Geoff Marcy, Doug Lin and Steven
Lubow for helpful and informative discussions.  We also thank an
anonymous referee for suggestions which helped us to improve the
paper.  We acknowledge support by UC Berkeley's Center for
Integrative Planetary Science (EWT) and NASA's Origins of Solar Systems
program through Grants NAG5-8861 (EWT) and 188-07-21-03-00 (JJL).

\bibliography{astrorefs}

\begin{thebibliography}{45}
\expandafter\ifx\csname natexlab\endcsname\relax\def\natexlab#1{#1}\fi

\bibitem[{{Benedict} {et~al.}(2002){Benedict}, {McArthur}, {Forveille},
  {Delfosse}, {Nelan}, {Butler}, {Spiesman}, {Marcy}, {Goldman}, {Perrier},
  {Jefferys}, \& {Mayor}}]{2002ApJ...581L.115B}
{Benedict}, G.~F., {McArthur}, B.~E., {Forveille}, T., {Delfosse}, X.,
  {Nelan}, E., {Butler}, R.~P., {Spiesman}, W., {Marcy}, G., {Goldman}, B.,
  {Perrier}, C., {Jefferys}, W.~H., \& {Mayor}, M. 2002, \apjl, 581, L115

\bibitem[{{Bodenheimer} {et~al.}(2000){Bodenheimer}, {Hubickyj}, \&
  {Lissauer}}]{2000Icar..143....2B}
{Bodenheimer}, P., {Hubickyj}, O., \& {Lissauer}, J.~J. 2000, Icarus, 143, 2

\bibitem[{{Bryden} {et~al.}(2000){Bryden}, {R{\' o}{\. z}yczka}, {Lin}, \&
  {Bodenheimer}}]{2000ApJ...540.1091B}
{Bryden}, G., {R{\' o}{\. z}yczka}, M., {Lin}, D.~N.~C., \& {Bodenheimer}, P.
  2000, \apj, 540, 1091

\bibitem[{{Cabot} {et~al.}(1987){Cabot}, {Canuto}, {Hubickyj}, \&
  {Pollack}}]{1987Icar...69..387C}
{Cabot}, W., {Canuto}, V.~M., {Hubickyj}, O., \& {Pollack}, J.~B. 1987, Icarus,
  69, 387

\bibitem[{{Chiang} \& {Goldreich}(1997)}]{1997ApJ...490..368C}
{Chiang}, E.~I., \& {Goldreich}, P. 1997, \apj, 490, 368

\bibitem[{{Cumming} {et~al.}(2003){Cumming}, {Marcy}, {Butler}, \&
  {Vogt}}]{2003asp...cumming}
{Cumming}, A., {Marcy}, G.~W., {Butler}, R.~P., \& {Vogt}, S.~S. 2003, in to
  appear in Scientific Frontiers in Research on Extrasolar Planets (ASP
  Conference Series)

\bibitem[{{Dubrulle}(1993)}]{1993Icar..106...59D}
{Dubrulle}, B. 1993, Icarus, 106, 59

\bibitem[{{Duncan} {et~al.}(1998){Duncan}, {Levison}, \&
  {Lee}}]{1998AJ....116.2067D}
{Duncan}, M.~J., {Levison}, H.~F., \& {Lee}, M.~H. 1998, \aj, 116, 2067

\bibitem[{{Gammie}(1996)}]{1996ApJ...457..355G}
{Gammie}, C.~F. 1996, \apj, 457, 355

\bibitem[{{Goldreich}(1965)}]{1965MNRAS.130..159G}
{Goldreich}, P. 1965, \mnras, 130, 159

\bibitem[{{Goldreich} \& {Sari}(2003)}]{2002...goldreich...sari}
{Goldreich}, P., \& {Sari}, R. 2003, ApJ, submitted

\bibitem[{{Gomes}(2000)}]{2000AJ....120.2695G}
{Gomes}, R.~S. 2000, \aj, 120, 2695

\bibitem[{{Haisch} {et~al.}(2001){Haisch}, {Lada}, \&
  {Lada}}]{2001ApJ...553L.153H}
{Haisch}, K.~E., {Lada}, E.~A., \& {Lada}, C.~J. 2001, \apjl, 553, L153

\bibitem[{{Hayashi}(1981)}]{1981PThPh..70...35H}
{Hayashi}, C. 1981, Prog.~Theor.~Phys., 70, 35

\bibitem[{{Kalas} \& {Jewitt}(1995)}]{1995AJ....110..794K}
{Kalas}, P., \& {Jewitt}, D. 1995, \aj, 110, 794

\bibitem[{{Kley}(2000)}]{2000MNRAS.313L..47K}
{Kley}, W. 2000, \mnras, 313, L47

\bibitem[{{Laughlin} \& {Chambers}(2001)}]{2001ApJ...551L.109L}
{Laughlin}, G., \& {Chambers}, J.~E. 2001, \apjl, 551, L109

\bibitem[{{Lee} \& {Peale}(2002)}]{2002ApJ...567..596L}
{Lee}, M.~H., \& {Peale}, S.~J. 2002, \apj, 567, 596

\bibitem[{{Lee} \& {Peale}(2003)}]{manhoi_dda_2003}
---. 2003, AAS/Division on Dynamical Astronomy Meeting \# 34

\bibitem[{{Levison} {et~al.}(1998){Levison}, {Lissauer}, \&
  {Duncan}}]{1998AJ....116.1998L}
{Levison}, H.~F., {Lissauer}, J.~J., \& {Duncan}, M.~J. 1998, \aj, 116, 1998

\bibitem[{{Lin} {et~al.}(2000){Lin}, {Papaloizou}, {Terquem}, {Bryden}, \&
  {Ida}}]{2000prpl.conf.1111L}
{Lin}, D.~N.~C., {Papaloizou}, J.~C.~B., {Terquem}, C., {Bryden}, G., \& {Ida},
  S. 2000, Protostars and Planets IV, 1111

\bibitem[{{Lissauer}(1987)}]{1987pggp.rept..100L}
{Lissauer}, J.~J. 1987, in Planetary Geology and Geophysics Program Report,
  100--+

\bibitem[{{Lissauer} {et~al.}(1984){Lissauer}, {Peale}, \&
  {Cuzzi}}]{1984Icar...58..159L}
{Lissauer}, J.~J., {Peale}, S.~J., \& {Cuzzi}, J.~N. 1984, Icarus, 58, 159

\bibitem[{{Lubow} \& {Ogilvie}(2001)}]{2001ApJ...560..997L}
{Lubow}, S.~H., \& {Ogilvie}, G.~I. 2001, \apj, 560, 997

\bibitem[{{Malhotra}(1998)}]{1998LPI....29.1476M}
{Malhotra}, R. 1998, in 29th Annual Lunar and Planetary Science Conference,
  March 16-20, 1998, Houston, TX, Vol.~29, 1476

\bibitem[{{Malhotra} \& {Dermott}(1990)}]{1990Icar...85..444M}
{Malhotra}, R., \& {Dermott}, S.~F. 1990, Icarus, 85, 444

\bibitem[{{Marcy} {et~al.}(2001){Marcy}, {Butler}, {Fischer}, {Vogt},
  {Lissauer}, \& {Rivera}}]{2001ApJ...556..296M}
{Marcy}, G.~W., {Butler}, R.~P., {Fischer}, D., {Vogt}, S.~S., {Lissauer},
  J.~J., \& {Rivera}, E.~J. 2001, \apj, 556, 296

\bibitem[{{Masset} \& {Snellgrove}(2001)}]{2001MNRAS.320L..55M}
{Masset}, F., \& {Snellgrove}, M. 2001, \mnras, 320, L55

\bibitem[{{Mizuno} {et~al.}(1978){Mizuno}, {Nakazawa}, \&
  {Hayashi}}]{1978PThPh..60..699M}
{Mizuno}, H., {Nakazawa}, K., \& {Hayashi}, C. 1978, Prog.~Theor.~Phys., 60,
  699

\bibitem[{{Mouillet} {et~al.}(1997){Mouillet}, {Larwood}, {Papaloizou}, \&
  {Lagrange}}]{1997MNRAS.292..896M}
{Mouillet}, D., {Larwood}, J.~D., {Papaloizou}, J.~C.~B., \& {Lagrange}, A.~M.
  1997, \mnras, 292, 896

\bibitem[{{Murray} \& {Dermott}(1999)}]{1999..Murray..Dermott..book}
{Murray}, C.~D., \& {Dermott}, S.~F. 1999, Solar System Dynamics (Cambridge
  University Press)

\bibitem[{{Nelson} {et~al.}(2000){Nelson}, {Papaloizou}, {Masset}, \&
  {Kley}}]{2000MNRAS.318...18N}
{Nelson}, R.~P., {Papaloizou}, J.~C.~B., {Masset}, F., \& {Kley}, W. 2000,
  \mnras, 318, 18

\bibitem[{{Papaloizou} {et~al.}(2001){Papaloizou}, {Nelson}, \&
  {Masset}}]{2001A&A...366..263P}
{Papaloizou}, J.~C.~B., {Nelson}, R.~P., \& {Masset}, F. 2001, \aap, 366, 263

\bibitem[{{Peale}(1976)}]{1976ARA&A..14..215P}
{Peale}, S.~J. 1976, \araa, 14, 215

\bibitem[{{Pollack} {et~al.}(1996){Pollack}, {Hubickyj}, {Bodenheimer},
  {Lissauer}, {Podolak}, \& {Greenzweig}}]{1996Icar..124...62P}
{Pollack}, J.~B., {Hubickyj}, O., {Bodenheimer}, P., {Lissauer}, J.~J.,
  {Podolak}, M., \& {Greenzweig}, Y. 1996, Icarus, 124, 62

\bibitem[{{Rivera} \& {Lissauer}(2001)}]{2001ApJ...558..392R}
{Rivera}, E.~J., \& {Lissauer}, J.~J. 2001, \apj, 558, 392

\bibitem[{{Snellgrove} {et~al.}(2001){Snellgrove}, {Papaloizou}, \&
  {Nelson}}]{2001A&A...374.1092S}
{Snellgrove}, M.~D., {Papaloizou}, J.~C.~B., \& {Nelson}, R.~P. 2001, \aap,
  374, 1092

\bibitem[{{Strom} {et~al.}(1990){Strom}, {Edwards}, \&
  {Skrutskie}}]{1990csss....6..275S}
{Strom}, S.~E., {Edwards}, S., \& {Skrutskie}, M.~F. 1990, in Cool Stars,
  Stellar Systems, and the Sun; Proceedings of the 6th Cambridge Workshop,
  Seattle, WA, Sept. 18-21, 1989 (A91-44876 19-90). San Francisco, CA,
  Astronomical Society of the Pacific, Vol.~6, 275--288

\bibitem[{{Thommes} {et~al.}(2003){Thommes}, {Duncan}, \&
  {Levison}}]{2003Icarus...tdl...oligarchy}
{Thommes}, E.~W., {Duncan}, M.~J., \& {Levison}, H.~F. 2003, Icarus

\bibitem[{{Thommes} \& {Lissauer}(2003)}]{thommes_lissauer_03}
{Thommes}, E.~W., \& {Lissauer}, J.~J. 2003, in to appear in Proceedings of
  STSci Astrophsyics of Life Symposium, May 6-9, 2002 (ASP Conference Series)

\bibitem[{{Tittemore} \& {Wisdom}(1989)}]{1989Icar...78...63T}
{Tittemore}, W.~C., \& {Wisdom}, J. 1989, Icarus, 78, 63

\bibitem[{{Wahhaj} {et~al.}(2003){Wahhaj}, {Koerner}, {Ressler}, {Werner},
  {Backman}, \& {Sargent}}]{2003...wahhaj}
{Wahhaj}, Z., {Koerner}, D.~W., {Ressler}, M.~E., {Werner}, M.~W., {Backman},
  D.~E., \& {Sargent}, A.~I. 2003, ApJ, 584, L27

\bibitem[{{Ward}(1997)}]{1997Icar..126..261W}
{Ward}, W.~R. 1997, Icarus, 126, 261

\bibitem[{{Wisdom} \& {Holman}(1991)}]{1991AJ....102.1528W}
{Wisdom}, J., \& {Holman}, M. 1991, \aj, 102, 1528

\bibitem[{{Yu} \& {Tremaine}(2001)}]{2001AJ....121.1736Y}
{Yu}, Q., \& {Tremaine}, S. 2001, \aj, 121, 1736

\end{thebibliography}

\end{document}